\documentclass[conference]{IEEEtran}
\IEEEoverridecommandlockouts
\usepackage{cite}
\usepackage{amsmath,amssymb,amsfonts}
\usepackage{algorithmic}
\usepackage{graphicx}
\usepackage{float}
\usepackage{subfigure}
\usepackage{textcomp}
\usepackage{xcolor}
\usepackage{multirow}
\def\BibTeX{{\rm B\kern-.05em{\sc i\kern-.025em b}\kern-.08em
    T\kern-.1667em\lower.7ex\hbox{E}\kern-.125emX}}
\begin{document}

\title{VSCNN: Convolution Neural Network\\Accelerator With Vector Sparsity}

\author{\IEEEauthorblockN{Kuo-Wei, Chang, and Tian-Sheuan Chang}
\IEEEauthorblockA{\textit{Dept. of Electronics Engineering, 
 National Chiao Tung University
Hsinchu, Taiwan}
}
\thanks{K. Chang and T. Chang, "VSCNN: Convolution Neural Network Accelerator with Vector Sparsity," 2019 IEEE International Symposium on Circuits and Systems (ISCAS), 2019, pp. 1-5, doi: 10.1109/ISCAS.2019.8702471.}}

\maketitle

\begin{abstract}
Hardware accelerator for convolution neural network (CNNs) enables real time applications of artificial intelligence technology. However, most of the accelerators only support dense CNN computations or suffers complex control to support fine grained sparse networks. To solve above problem, this paper presents an efficient CNN accelerator with 1-D vector broadcasted input to support both dense network as well as vector sparse network with the same hardware and low overhead. The presented design achieves 1.93X speedup over the dense CNN computations.  

\end{abstract}


\section{Introduction}
Convolution neural networks (CNN) have been widely used in computer vision such as recognition\cite{1,2,3,4,5}, detection\cite{6,7,8,9,10}, and autonomous vehicles during recent years for its significant improvement over traditional approaches. However, computations of CNNs demands a lot of multiplications and accumulations (MACs), and millions of data amount per layer. Thus, hardware accelerators for CNNs are required to meet real time applications. 

Various hardware accelerators have been proposed recently\cite{11,12,13,14,15,16}, which can be divided into dense CNN or sparse CNN computation types. The dense CNN types \cite{11,12,13,14} assume continuous and regular computational data flow to the hardware accelerator, which results in simple and regular systolic array or filter-like architecture. However, CNN computations contain a lot of zeros in its weight and input activations due to model pruning \cite{17} and popular ReLU activation function. Exploring sparsity offers a significant speedup option of the hardware accelerator. But this type of sparsity is a fine grained sparse structure as shown in Fig. 1, where the zero distribution is abundant but irregular and bad for hardware design. To achieve sparse CNN computation, \cite{11} tried the gated input for zero input with the same dense CNN design to save power, which did not save computation cycles. \cite{15} skipped zero weight computation on its single instruction multiple data (SIMD) array by the zero weight indexing and their distance. \cite{16} explored both zero weight and input with a nonzero data indexing system, computed them by a 2D multiplier array, and accumulated those sparse outputs with the help of coordinate computation to sort these irregular output. All these designs\cite{11,15,16} are for the fine grained sparsity. The irregularity of the fine grained sparsity results in significant area cost on the indexing system and data routing.

To reduce above cost while still explore the benefit of sparsity, this papers proposes a CNN accelerator to support dense CNN computation as well as vector sparse CNN on both weight and input. The vector sparsity as shown in Fig. 2 \cite{18} has vectors of zeros instead of fine grained ones, which enables regular hardware design and still offers zero skipping benefits as shown in our experimental results. To support this, the proposed design operates on a 1-D to 1-D matrix multiplication with broadcasted 1-D weight and input, which enables zero vector skipping easily. 

This CNN accelerator design has the following contributions
\begin{itemize}
	\item Support dense CNN, and vector sparse CNN in one design with the same accumulator flow and an index system for vector sparsity.
	\item Skip both zero weight data and input data to get great performance.
\end{itemize}
\begin{figure}[htbp]
	\centering
	\begin{minipage}[H]{4cm}
		\centering
		\includegraphics[height=25mm]{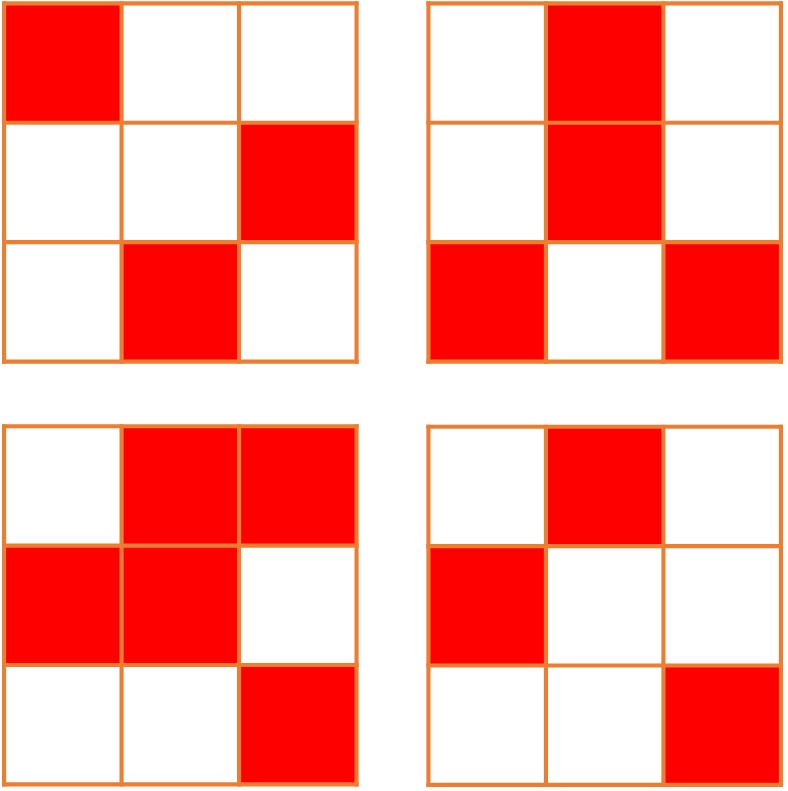}
		\caption{Fine grained sparse structure}
	\end{minipage}
	\begin{minipage}[H]{4cm}
		\centering
		\includegraphics[height=25mm]{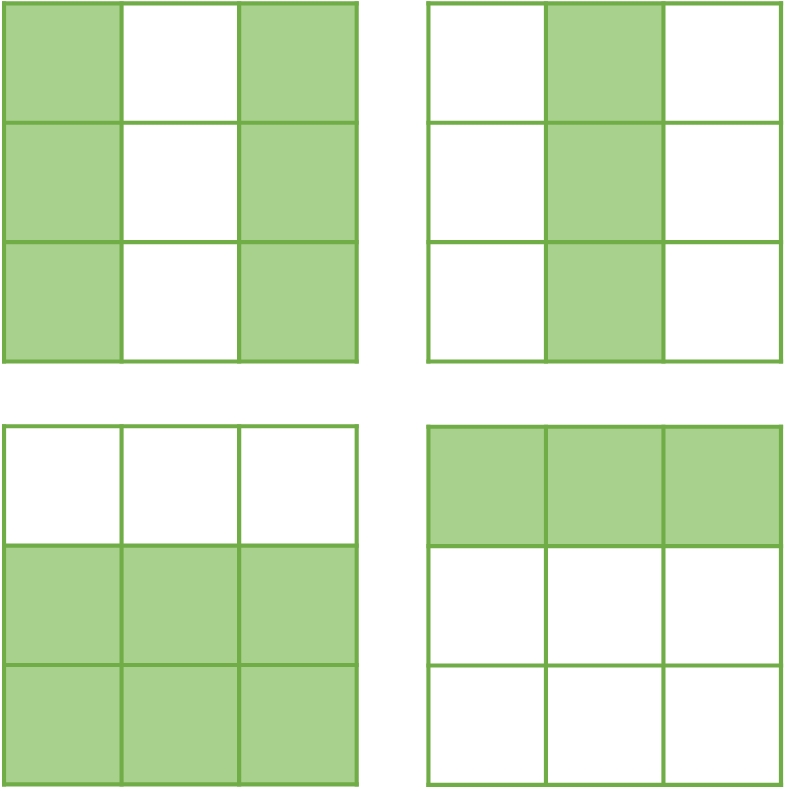}
		\caption{Vector sparse structure}
	\end{minipage}
\end{figure}

\begin{figure}[H]
	\centering{\includegraphics[height=48mm]{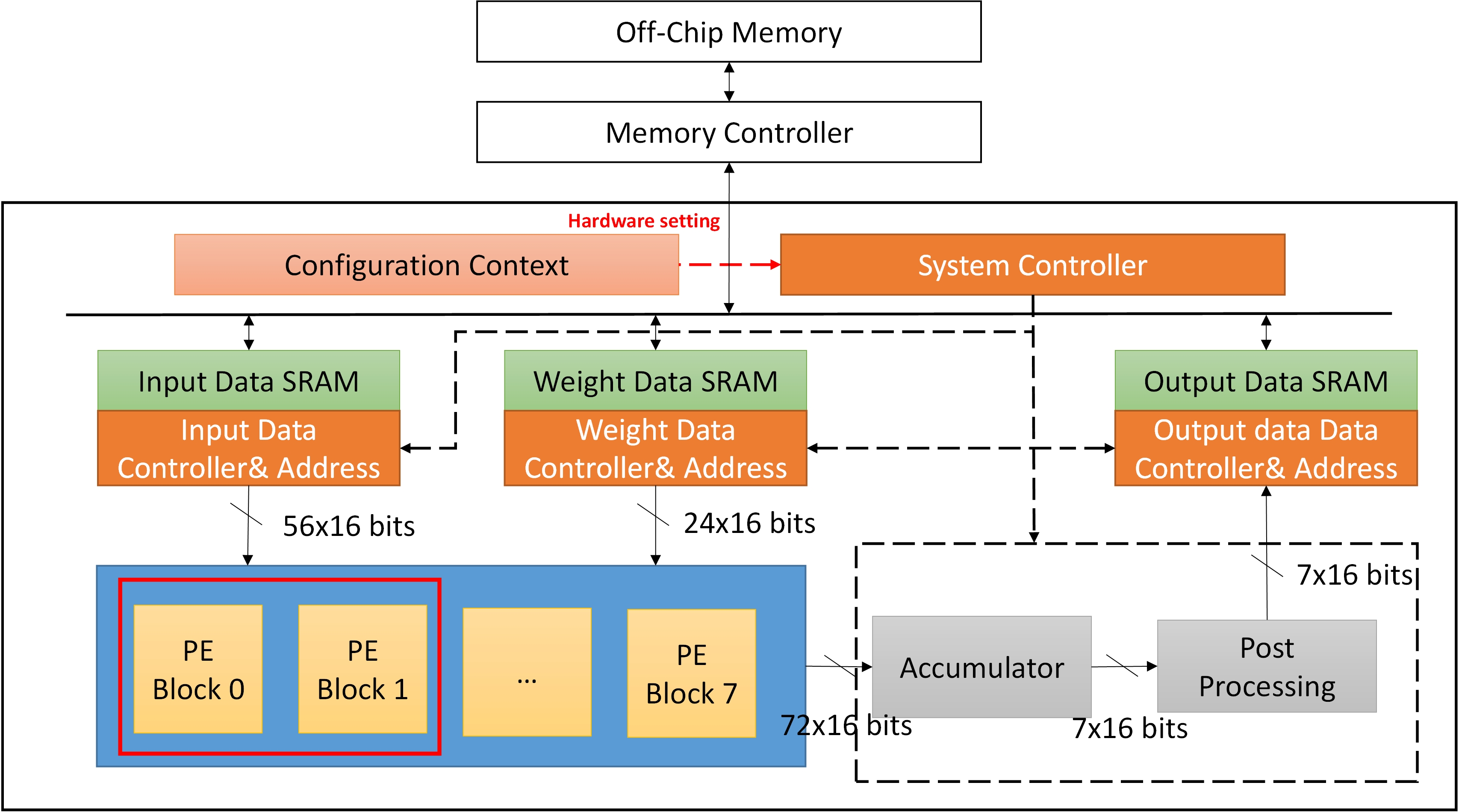}}
	\caption{The proposed system architecture}
	\label{fig:Top}
\end{figure}

The rest of the paper is organized as following. Section II shows the overview of our proposed architecture. Section III gives detailed data flow for dense CNN and vector sparse CNN. The implementation results and comparison are shown in section IV. Finally, we conclude in section V.

\section{Architecture}
\subsection{Overview}
Fig.~\ref{fig:Top} shows the proposed system architecture. This design first gets the input and weight data from external memory and stores them into the local SRAM buffers for the following repeated access. These data are fed into the processing element (PE) array to compute convolution and then accumulated through the accumulator according to the index. During the accumulation, these partial sum of the convolution results are stored in a local SRAM buffer to avoid unnecessary external memory access until the final accumulated output is generated. These accumulated output is processed by the post processing unit for following activation functions, normalization, and zero detection.  The final output which are non-zero vector will be sent back to external DRAM. The whole process is controlled by the system controller according to the configuration context. The data access of input, weight and output are controlled by their SRAM buffer controllers to sequentially accessing the data.
\begin{figure}[t]
	\centering{\includegraphics[height=60mm]{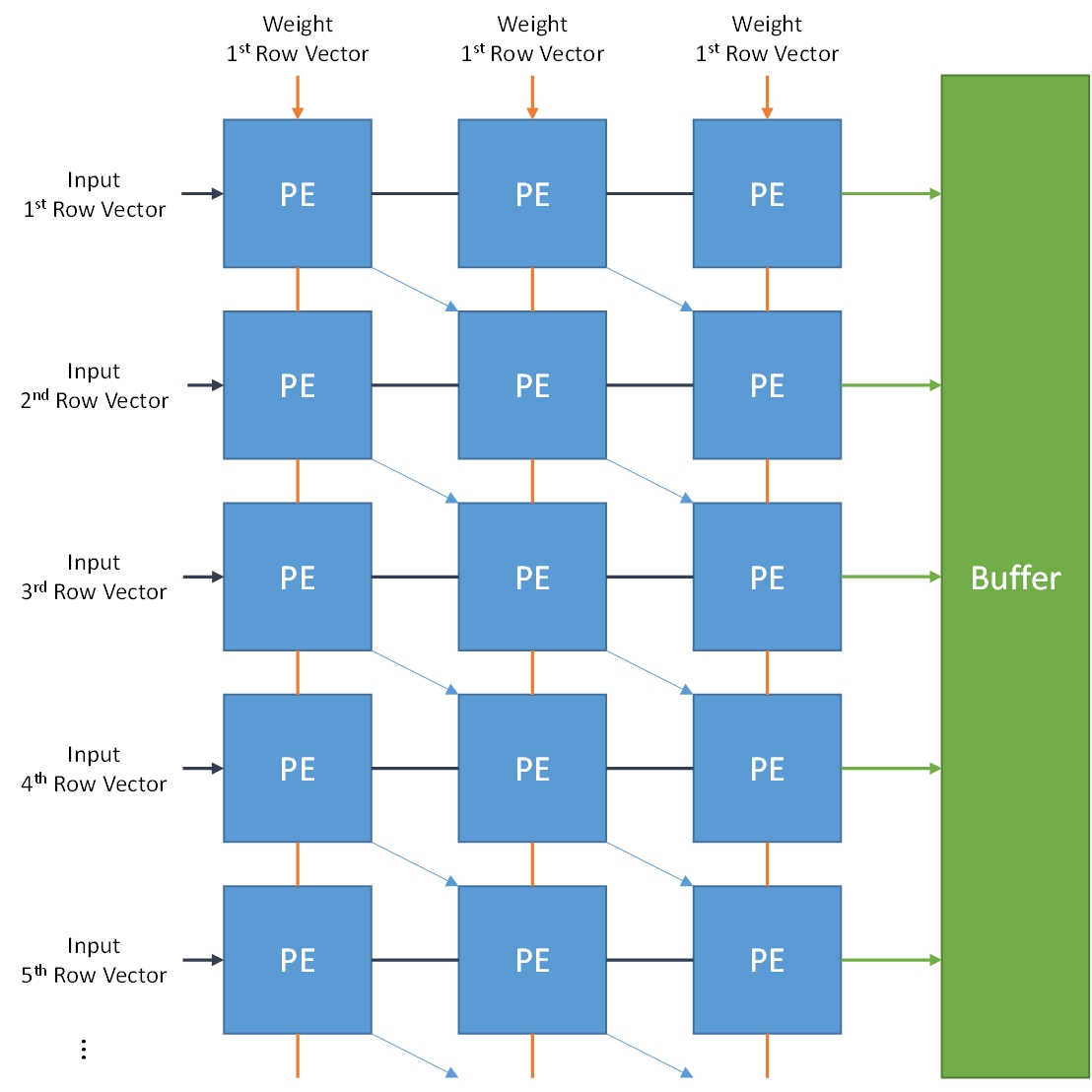}}
	\caption{The PE array architecture}
	\label{fig:overview_pe}
\end{figure}
\begin{figure}[t]
	\centering{\includegraphics[height=60mm]{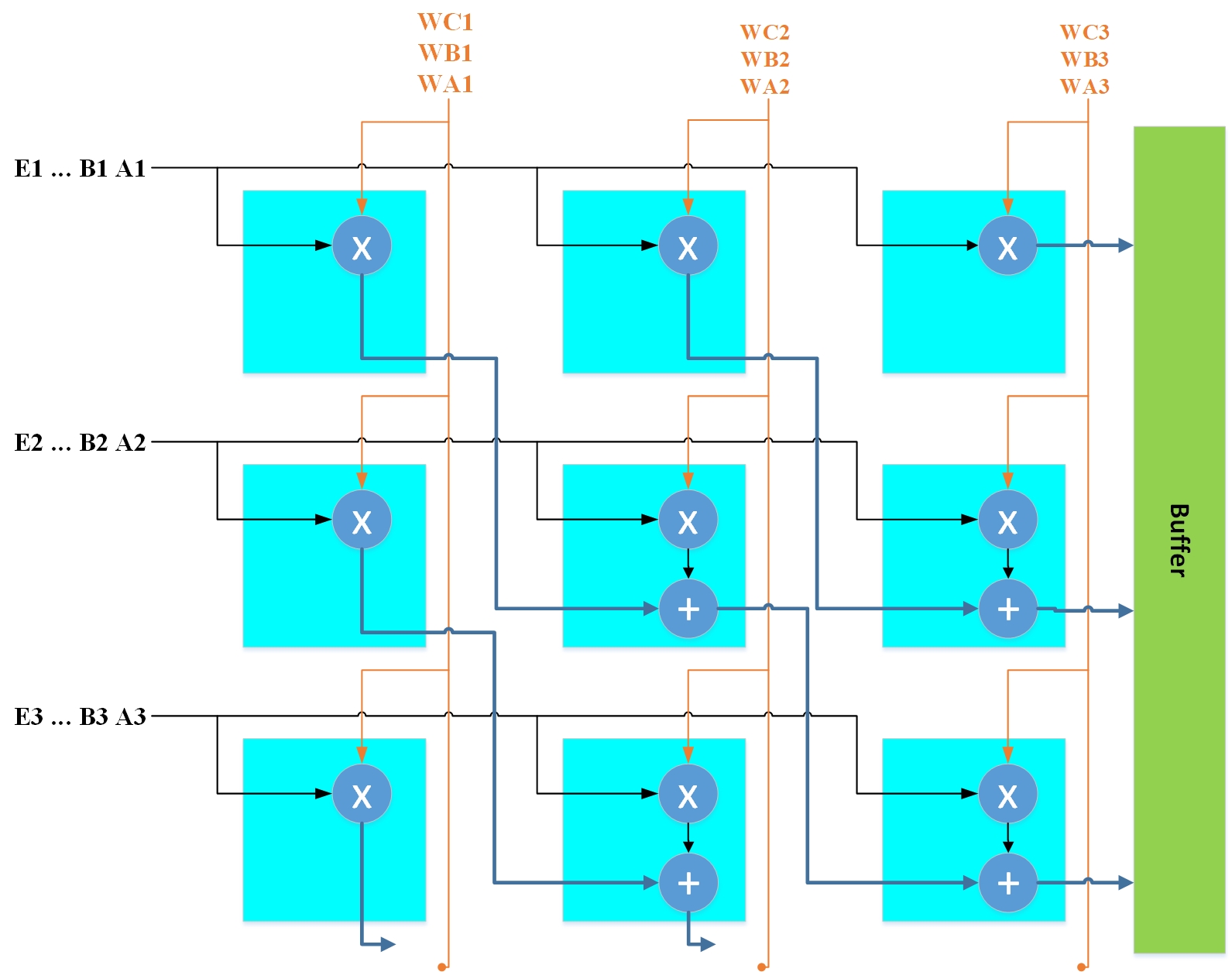}}
	\caption{The overview of processing element}
	\label{fig:detail_pe}
\end{figure}

\subsection{PE array}
The PE array is the main processing core of this proposed design as shown in Fig.~\ref{fig:overview_pe} and Fig.~\ref{fig:detail_pe}. Each PE as in Fig.~\ref{fig:detail_pe} contains one multiplier to multiply input and weight, and adder 
\begin{figure}[H]
	\centering{\includegraphics[height=30mm]{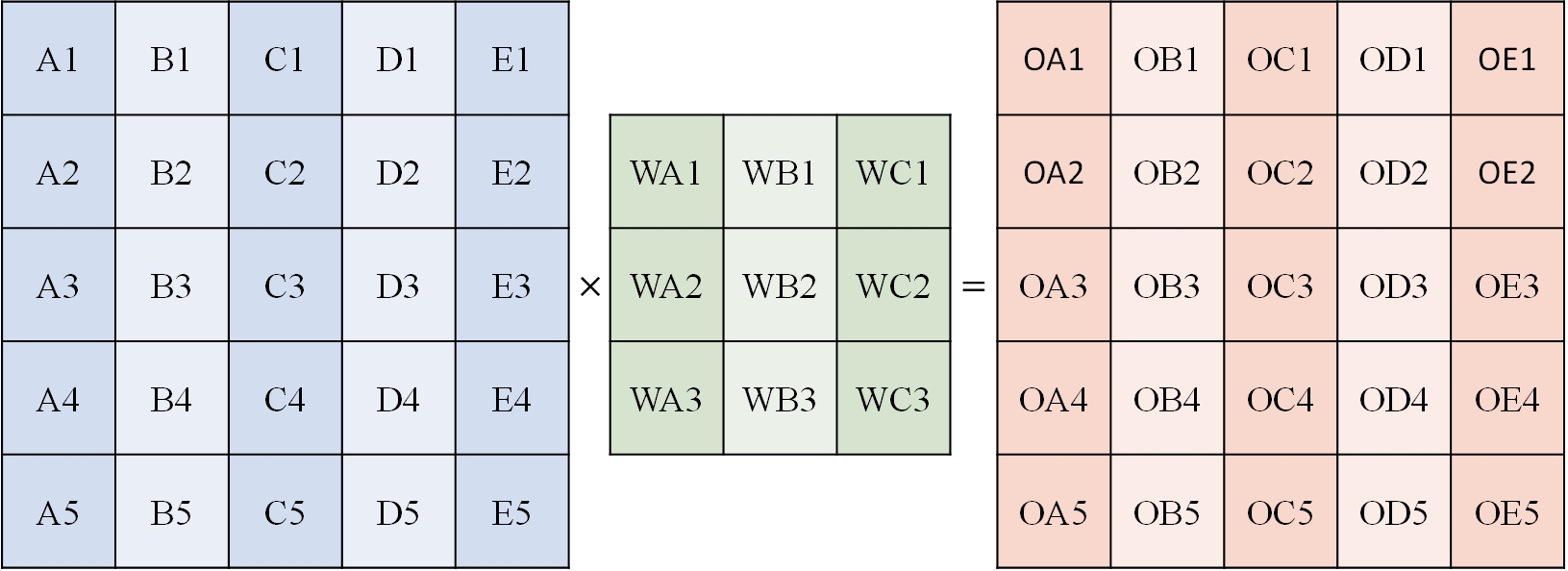}}
	\caption{An convolution example with 5x5 input with padding 1 and 3x3 weight to generate 5x5 output}
	\label{fig.example}
\end{figure}
for partial sum accumulation. In this design, the weights are not stored locally in each PE as in other designs since such design style is not suitable for sparse computation. Instead, the input data are broadcasted horizontally and the weight data are broadcasted vertically to support both dense and vector sparse computation. The partial results are then propagated diagonally along the PE array, as shown in Fig.~\ref{fig:detail_pe} to accumulate in the same cycle. 

The hardware utilization of the PE array depends on how to map the kernels and inputs to the array and the sparsity of the network. Since the 3x3 convolution is the most widely filter in current CNNs, our architecture has optimized the convolution process for 3x3 filters with the unit stride for full hardware utilization. Other filter sizes and non-unit strides can be supported as well by a suitable mapping method \cite{13}.

\begin{figure}[b]
	\centering{\includegraphics[height=60mm]{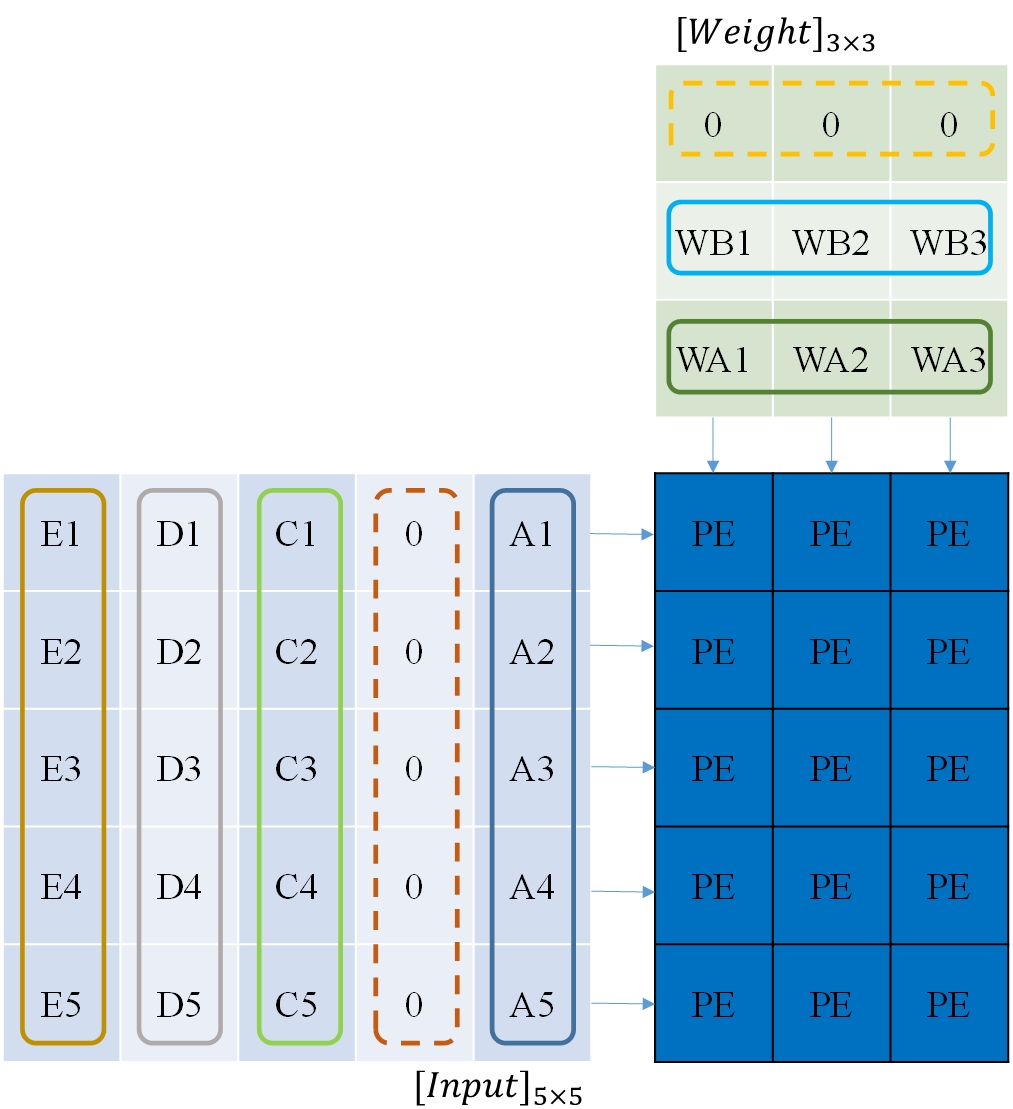}}
	\caption{Illustration of data flow
		(dashed line block represents all zero vector in sparse CNN)}
	\label{fig.data_flow}
\end{figure}

\begin{table*}[t]
	\centering
	\caption{Timing Diagram}
	\begin{tabular}{|c||c|c|c|c|c|c|c|c|c|c|}
		\hline
		\multicolumn{11}{|c|}{Dense CNN Timing Diagram}\\
		\hline
		Cycle &1 &2 &3 &4 &5 &6&7&8&9 & ...\\
		\hline
		
		Input &\multicolumn{3}{|c|}{A1-A5} &\multicolumn{3}{|c|}{B1-B5}& \multicolumn{3}{|c|}{C1-C5}& ...\\
		\hline
		Weight&WA1-WA3 &WB1-WB3 &WC1-WC3 &WA1-WA3 &WB1-WB3 &WC1-WC3 & WA1-WA3 &WB1-WB3 &WC1-WC3&  ...\\
		\hline
		Output & OB1-OB5& OA1-OA5&x & OC1-OC5&OB1-OB5 &OA1-OA5 & OD1-OD5& OC1-OC5&OB1-OB5& ... \\		
		\hline
		\multicolumn{11}{|c|}{Sparse CNN Timing Diagram}\\
		\hline
		Cycle &1 &2 &3 &4 &5 &6&7&8&\multicolumn{2}{|c|}{...}\\
		\hline
		
		Input &\multicolumn{2}{|c|}{A1-A5} &\multicolumn{2}{|c|}{C1-C5}& \multicolumn{2}{|c|}{D1-D5}&\multicolumn{2}{|c|}{E1-E5}&\multicolumn{2}{|c|}{...}\\
		\hline
		Weight&WA1-WA3 &WB1-WB3  &WA1-WA3 &WB1-WB3  & WA1-WA3 &WB1-WB3& WA1-WA3 &WB1-WB3&\multicolumn{2}{|c|}{...}\\
		\hline
		Output & OB1-OB5& OA1-OA5 & OC1-OC5&OB1-OB5  & OE1-OE5& OD1-OD55&x&OE1-E5&\multicolumn{2}{|c|}{...} \\		
		\hline
	\end{tabular}
	\label{time_dia}
\end{table*}

\begin{figure*}[t]
	\centering{\includegraphics[height=75mm]{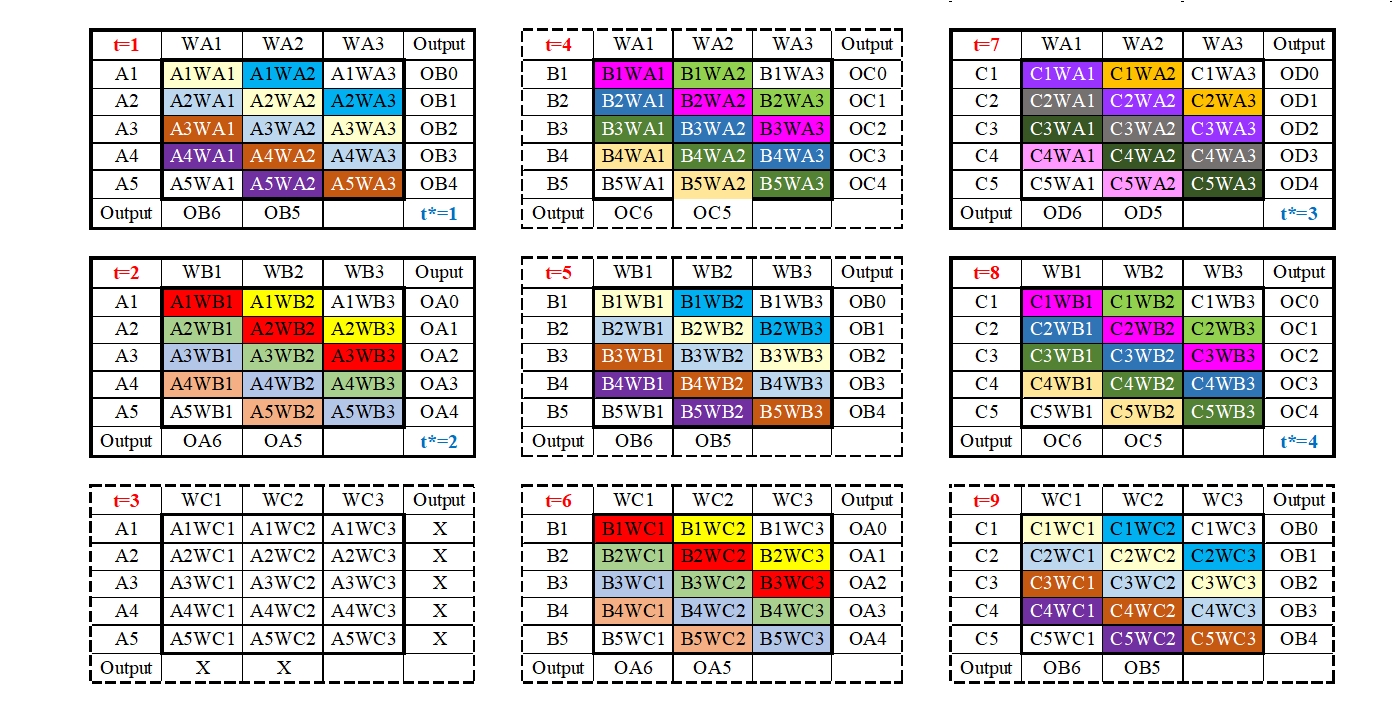}}
	\caption{Dataflow chart for dense and sparse CNN computation in Fig. 8, where the dashed line blocks will be skipped at the sparse CNN case. In each block, the element with the same color will be summed together in a PE. OA0, OA6, OB0, OB6, … are for zero padding boundary computation.   t* represents cycles for sparse CNN.}
	\label{fig.total_data_flow}
\end{figure*}

 \begin{figure}[t]
	\centering{\includegraphics[height=50mm]{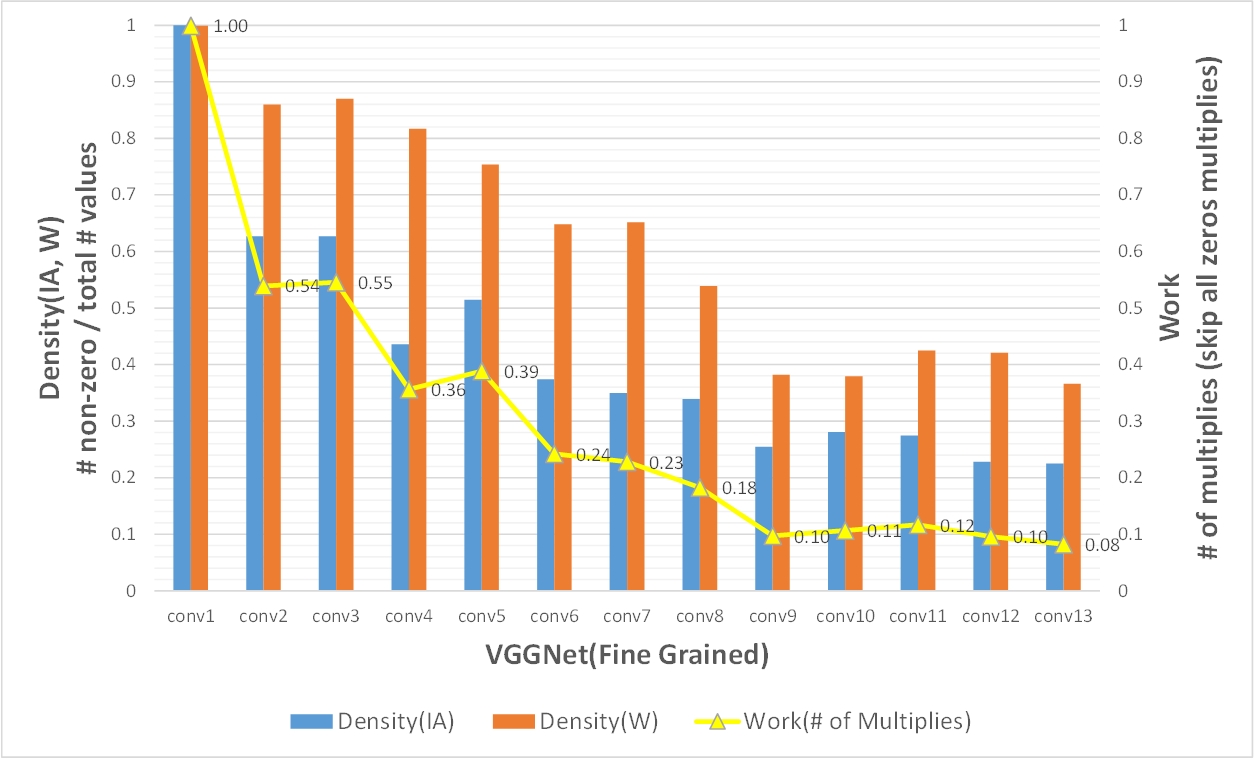}}
	\caption{Density ratio of input and weight and work with fine grained sparse}
	\label{fig:sparse_ratio}
\end{figure}

\section{Dense and Sparse Data Flow}
Fig.~\ref{fig.data_flow} shows the data flow of the CNN computation with 15 PEs for a 5x5 input with padding 1, and 3x3 filer kernel example as in Fig.~\ref{fig.example}. At the first cycle, the first column of the weight filter, WA1 to WA3, is broadcasted vertically to the PE array. The corresponding first column of the input activations, A1 to A5, is also broadcasted horizontally to the PE array. The vector to vector multiplication results are also summed together along the diagonal direction at the same cycle to generate part of the results of OB1 to OB5, as illustrated in Table.~\ref{time_dia} and Fig.~\ref{fig.total_data_flow}. These partial results (e.g. OB1 to OB5 at t = 1) are stored in the buffer and accumulated with the next partial results with the same index (e.g. OB1 to OB5 at t = 5, 9).

For dense CNN computation as in Fig.~\ref{fig.total_data_flow}, each output will take 3 different cycles for 3x3 filters, and 15 cycles for 5x5 input. For sparse CNN computation, zero input data and weight data as denoted by the dashed line block will not be in SRAM, so they will be skipped and not be computed as shown in Table.~\ref{time_dia} and Fig.~\ref{fig.total_data_flow}. Only the nonzero part will be in SRAM and sent to the PE array and accumulated with the same index system. Thus, the computation cycles are reduced (e.g. t=3, 4, 5, 6 and 9 in Fig.~\ref{fig.total_data_flow}) while still keep the computation regular. As result, we only need 8 cycles for this sparse CNN, saving 47\% of cycles.

\begin{figure}[htbp]
	\centering{\includegraphics[height=50mm]{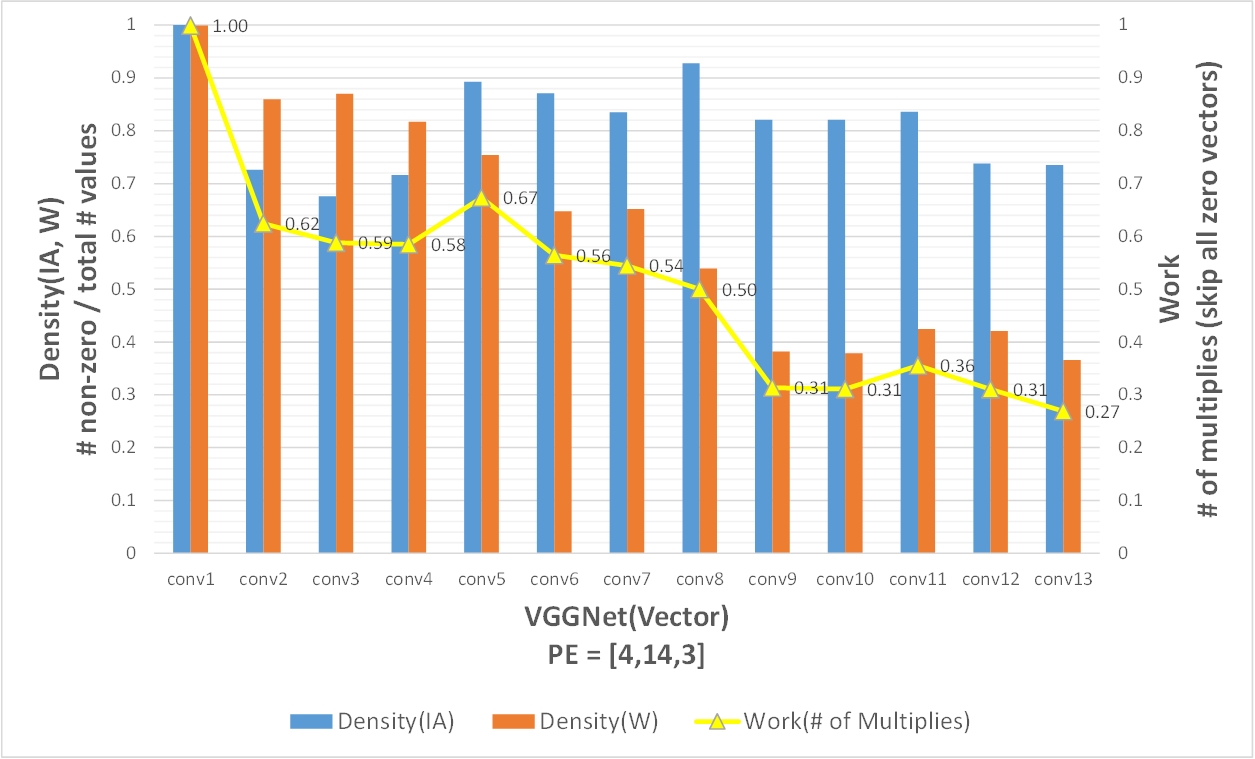}}
	\caption{Density ratio of input and weight and work with PE array 4 blocks, 14 rows, 3 columns}
	\label{fig:v_s_ratio14}
\end{figure}

\begin{figure}[htbp]
	\centering{\includegraphics[height=50mm]{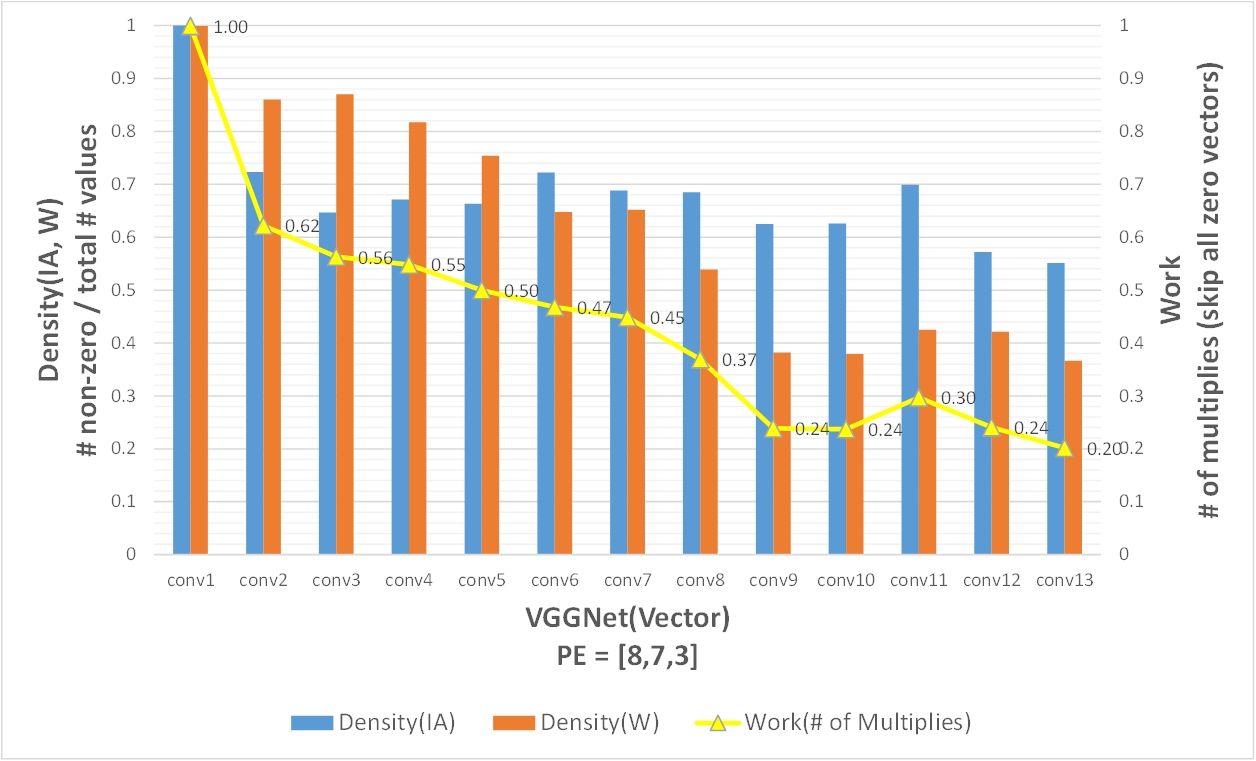}}
	\caption{Density ratio of input and weight and work with PE array 8 blocks, 7 rows, 3 columns}
	\label{fig:v_s_ratio7}
\end{figure}

\begin{figure}[htbp]
	\centering{\includegraphics[height=50mm]{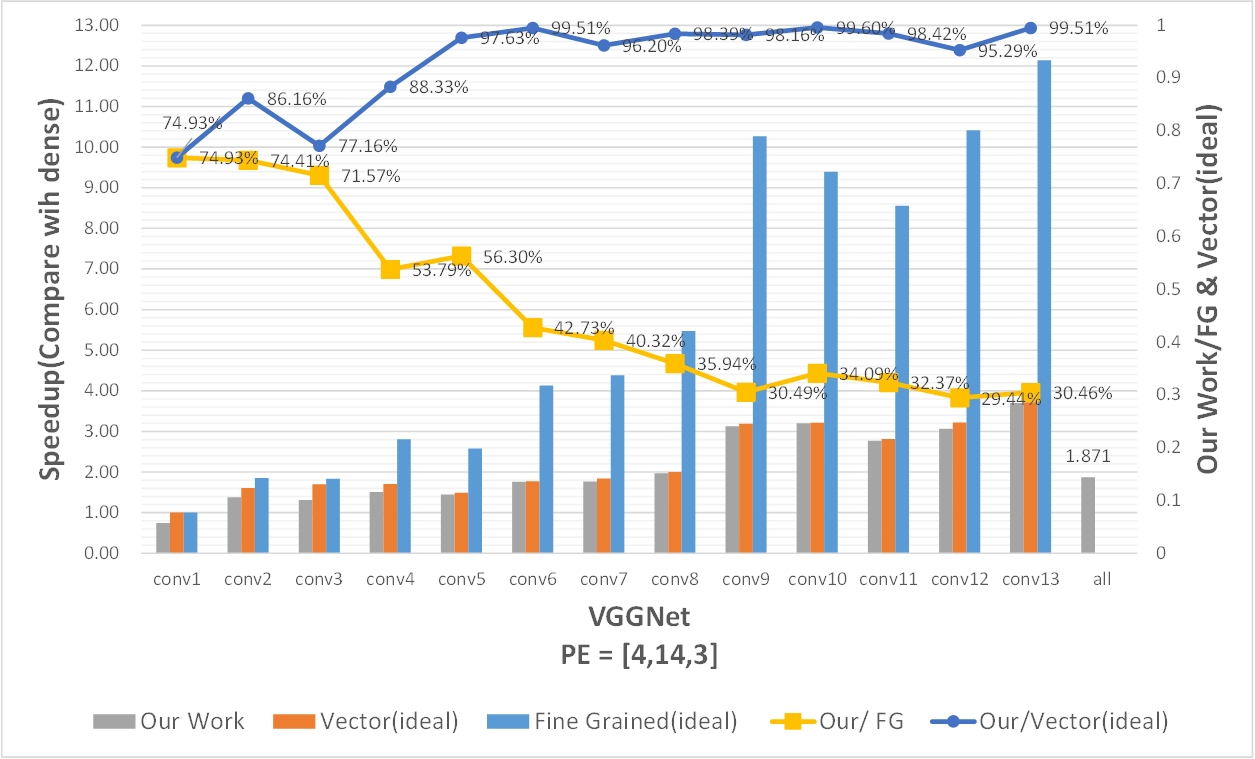}}
	\caption{Speedup of our work and ideal vector sparse and fine grained sparse network with PE array 4 blocks, 14 rows, 3 columns.}
	\label{fig:speedup_14}
\end{figure}

\begin{figure}[htbp]
	\centering{\includegraphics[height=50mm]{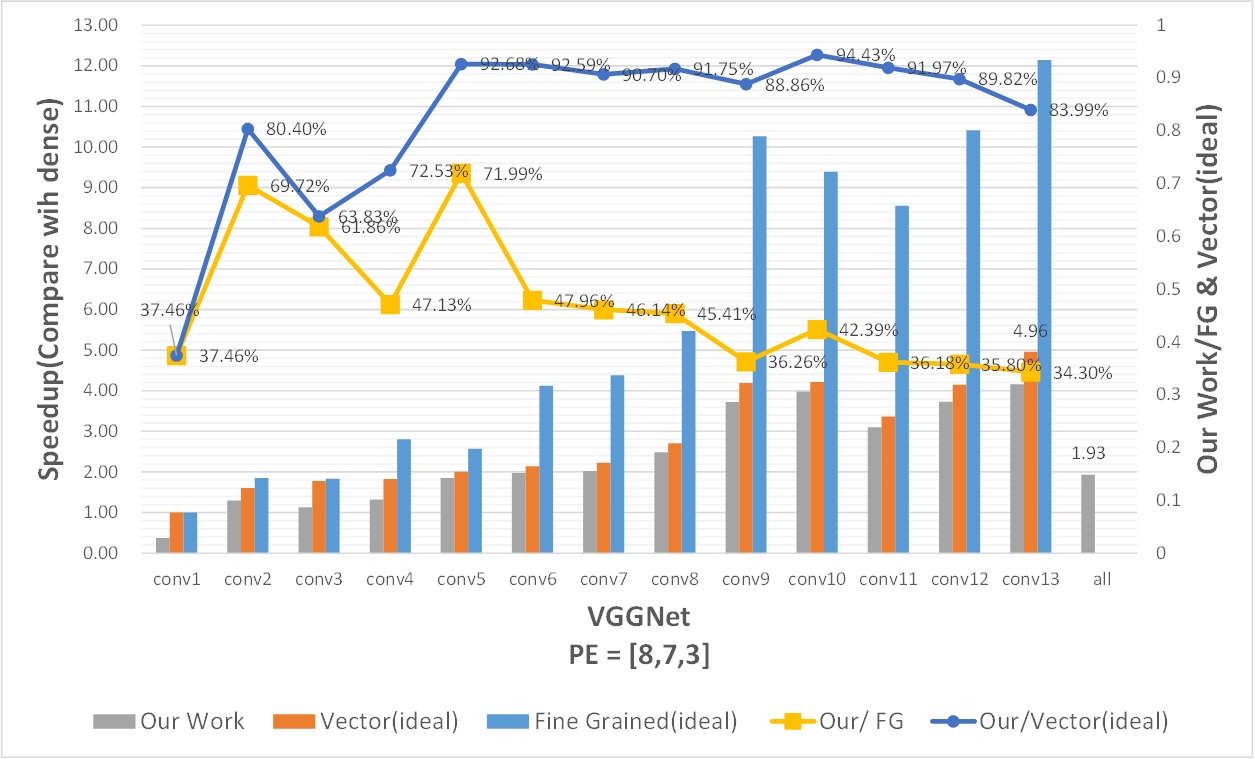}}
	\caption{Speedup of our work and ideal vector sparse and fine grained sparse network with PE array 8 blocks, 7 rows, 3 columns.}
	\label{fig:speedup_7}
\end{figure}

\section{Experimental Result}
The proposed architecture has been implemented and simulated with the VGG-16 model pretrained on the ImageNet dataset and pruned with the vector pruning method as \cite{18}. The accuracy only drop 0.08\% with density 23.5\%. The PE number used in the simulation is 168, arranged in two configurations: [4, 14, 3] 4 PE arrays, and 14 rows and 3 columns per PE array, [8, 7, 3] 8 PE arrays, and 7 row and 3columns per PE array. Such configurations are chosen to maximize the hardware utilization for above VGG-16 model execution. With above configuration, the input activation vector size is set to 14 or 7. Following output zero detection in post processing element and weight pruning, Fig.~\ref{fig:sparse_ratio} and Fig.~\ref{fig:v_s_ratio14} shows the nonzero data density of input activation and weight for fine grained sparsity and vector sparsity, respectively. As expected, the fine grained sparsity has lower density than that in the vector sparsity case. 

The speedup results of the proposed design are shown in Fig.~\ref{fig:speedup_14} and Fig.~\ref{fig:speedup_7}. When compared with the dense CNN computation, we can achieve 1.871X and 1.93X speedup for [4, 14, 3] and [8, 7, 3] cases, respectively. Small number of PE rows in the [8, 7, 3] case results in more zero vectors to skip, and thus higher speedup, but the difference is small. The zero computation that this design can skip is 92\% ([4, 14, 3] case) and 85\% ([8, 7, 3]) compared to their respective ideal vector sparse computation, and 46.6\% ([4, 14, 3] case) and 47.1\% ([8, 7, 3]) compared to the ideal fine grained computation. Our design is efficient to exploit almost all zero vectors. Small zero vector enables more zero skipping. 

When compared to the fine grained design in \cite{16}, our design overhead is very small compared to the complex index, accumulator and routing in \cite{16}. The speedup over the dense CNN in \cite{16} is about 3X, which roughly exploits 66\% of ideal fine grained zero computation. In comparison, our design can exploits 47\% in average of ideal fine grained zero computation 
to achieve 1.93X speedup with small area overhead. Our design is more hardware efficient than the previous design.

\section{Conclusion}
This paper proposes a CNN hardware accelerator that can support dense CNN and vector sparse CNN incurring small area overhead with broadcasted 1-D input activation vector and 1-D weight vector data flow. This design can achieve 1.93X speedup over the dense CNN computation by exploiting around 90\% of the ideal zero vector computation. 

\section*{Acknowledgment}
This work was supported by Ministry of Science and Technology, Taiwan, under Grant 108-2634-F-009 -005, 107-2119-M-009-019, and Research of Excellence program 106‐2633‐E‐009‐001.

\end{document}